\documentstyle[pra,aps,preprint]{revtex}

\begin{document}

\tightenlines

\draft
\title{An optically driven quantum dot quantum computer}
\author{G. D. Sanders, K. W. Kim, and W. C. Holton}
\address{Department of Electrical and Computer Engineering,
North Carolina State University\\
Raleigh, North Carolina 27695-7911}

\maketitle
\begin{abstract}

We propose a quantum computer structure based on coupled asymmetric 
single-electron quantum dots. Adjacent dots are strongly coupled by means 
of electric dipole-dipole interactions enabling rapid computation rates.  
Further, the asymmetric structures can be tailored for a long coherence
time. The result maximizes the number of computation cycles prior to loss
of coherence.

\end{abstract}

\pacs{PACS Number(s): 03.67.Lx, 73.20.Dx, 85.30.Vw}


The possibility that a computer could be built employing the laws of quantum 
physics has stimulated considerable interest in searching for useful
algorithms and a realizable physical implementation. Two useful
algorithms, exhaustive search \cite{bib:Grover97} and factorization
\cite{bib:Shor94}, have been 
discovered; others have been shown possible. Various approaches have been 
explored for possible physical implementations, including trapped 
ions \cite{bib:Cirac95}, cavity quantum electrodynamics
\cite{bib:Pellizzari95}, ensemble nuclear magnetic resonance
\cite{bib:NMR}, small Josephson junctions \cite{bib:Shnirman97},
optical devices incorporating beam splitters and phase shifters
\cite{bib:Cerf98}, and a number of solid state systems based on quantum dots
\cite{bib:BarencoDeutsch95,bib:Wang,bib:Kane,bib:Tanamot099,bib:Loss98}.
There are many advantages to quantum computing; however, the requirements
for such computers are very stringent, perhaps especially so for solid state
systems. Nevertheless, solid state quantum computers are very appealing
relative to other proposed physical implementations. For example,
semiconductor-manufacturing technology is immediately applicable to the
production of quantum computers of the proper implementation that is
readily scalable due to its artificially fabricated nature.

In this paper, we propose a manufactured solid state implementation based 
on advanced nanotechnology that seems capable of physical implementation.
It consists of an ensemble of "identical" semiconductor pillars, each 
consisting of a vertical stack of coupled asymmetric GaAs/AlGaAs 
single-electron quantum dots of differing sizes and material compositions 
so that each dot possesses a distinct energy structure. Qubit registers are 
based on the ground and first excited states of a single electron within 
each quantum dot. The asymmetric dots produce large built-in electrostatic 
dipole moments between the ground and excited states, and electrons in 
adjacent dots are coupled through an electric dipole-dipole interaction.
The dipole-dipole coupling between electrons in nonadjacent dots is less
by ten times the coupling between adjacent dots. Parameters of the
structure can be chosen to produce a well-resolved spectrum of
distinguishable qubits with adjacent qubits strongly coupled.
 The resulting ensemble of quantum computers may also be tuned electrically 
through metal interconnect to produce "identical" pillars.
In addition, the asymmetric potential can be designed so that dephasing
due to electron-phonon scattering and spontaneous emission is minimized.
The combination of strong dipole-dipole coupling and long dephasing times
make it possible to perform many computational steps.
Quantum computations may be carried out in complete analogy with the
operation of a NMR quantum computer, including the application of
refocusing pulses to decouple qubits not involved with a current
step in the computational process \cite{bib:Cory98}.
Final readout of the amplitude and phase
of the qubit states can be achieved through quantum state holography.
Amplitude and phase information are extracted through mixing the final
state with a reference state generated in the same system by an
additional delayed laser pulse and detecting the total time- and 
frequency- integrated fluorescence as a function of the delay
\cite{bib:Leichtle,bib:Weinacht}. Means of characterizing
the required laser pulses are described in Ref. \onlinecite{bib:Trebino}.

Our quantum register is similar to the n-type single-electron transistor
structure recently reported by Tarucha et al. \cite{bib:Tarucha96}. In
Tarucha's structure, source and drain are at the top and bottom of a
free standing pillar with a quantum well in the middle and a cylindrical
gate wrapped around the center of the pillar. In our design, a stacked
series of asymmetric GaAs/AlGaAs quantum wells are arrayed along the
pillar axis by first epitaxially growing planar quantum wells in a 
manner similar to that employed to produce surface emitting lasers
\cite{bib:Zhang}. By applying a
negative gate bias that depletes carriers near the surface, a parabolic
electrostatic potential is formed which provides confinement in the
radial direction. In the strong depletion regime, the curvature of the
parabolic radial potential is a function of the doping concentration.
To facilitate coupling to the laser field, the gate is made transparent 
using a reverse damascene process. The simultaneous
insertion of a single electron in each dot is accomplished by lining up the 
quantum dot ground state levels so they lie close to the Fermi level; 
a single electron is confined in each dot over a finite range of the gate 
voltage due to shell filling effects \cite{bib:Tarucha96}.
Strong electrostatic confinement in the radial direction serves to
keep the quantum dot electrons from interacting with the gate electrode,
phonon surface modes, localized surface impurities, and interface roughness
fluctuations. The electrostatic potential near the pillar axis is smooth in
the presence of small fluctuations in the pillar radius. By tuning the 
gate voltage, it is anticipated that size fluctuations between different
pillars can be compensated for.

In order to derive the structure 
parameters and estimate the dependence of the functional performance of 
this device, we assume that the quantum dot electron potential, V(r), can 
be expressed in cylindrical coordinates
as $V(\vec{r}) = V(z) + V(\rho)$, where $V(\rho)$ is a radial potential 
and $V(z)$ is the potential along the growth direction. This
separable potential assumption is a good approximation in the strong
depletion regime where only a single electron resides in each dot.
The assumption of a separable potential is commonly used in the study
of quantum dot structures and enables us to consider the $z$ and
$\rho$ motions separately \cite{bib:Tarucha96,bib:Jacak98}. The z-directional 
potential $V(z)$, shown schematically in the inset of Fig.\ \ref{density},
is a step potential formed by a layer of
$Al_{x}Ga_{1-x}As$ of thickness $B$ ($0 < z < B$) and a layer of
$GaAs$ of thickness $L - B$ ($B < z < L$). The resulting asymmetric quantum
dot/well is confined by $Al_{y}Ga_{1-y}As$ barriers with $y > x$.
The asymmetry of this structure is parameterized by the ratio $B/L$ where
$0 < B/L < 1$.

In the effective mass approximation, the qubit wavefunctions are
$\arrowvert i \rangle = R(\rho) \ \psi_{i}(z) \ u_{s}(\vec{r})$ ($i = 0,1$).
Here $R(\rho)$ is the ground state of the radial envelope function,
$\psi_{i}(z)$ is the envelope function along $z$,
and $u_{s}(\vec{r})$ is the $s$-like zone center Bloch function including
electron spin. For simplicity, we assume complete confinement by the
$Al_{y}Ga_{1-y}As$ barriers along the z direction. Then, the envelope 
function $\psi_{i}(z)$ is obtained by solving the time-independent 
Schr\"{o}dinger equation subject to the boundary conditions $\psi_{i}(0) 
= \psi_{i}(L) = 0$.  The energies of the qubit wavefunctions are given by
$E = E_{\rho} + E_{i}$ where $E_{\rho}$ is the energy associated with
$R(\rho)$ and $E_{i}$ is the energy associated with $\psi_{i}(z)$.
Since the present study primarily concerns coupling along the growth 
direction, analyses are conducted only in this direction.

Figure \ref{density} shows the probability density,
$\arrowvert \psi_{i}(z) \arrowvert ^{2}$,
as a function of position, $z$, for the two qubit states
$\arrowvert 0 \rangle$ and
$\arrowvert 1 \rangle$ in a $20 \ nm$ $GaAs/Al_{0.3}Ga_{0.7}As$
asymmetric quantum dot. The barrier thickness $B = 15 \ nm$ and the 
overall length of the dot is $L = 20 \ nm$. By choosing $B/L = 0.75$ and
$x = 0.3$, it is found that the ground state wavefunction
$\arrowvert 0 \rangle$ is strongly localized in the $GaAs$ region while the
$\arrowvert 1 \rangle$ wavefunction is strongly localized in the
$Al_{0.3}Ga_{0.7}As$ barrier. By appropriately choosing the
asymmetric quantum dot parameters, the qubit
wavefunctions can be spatially separated and a large difference
in the electrostatic dipole moments can be achieved.

The transition energy $\Delta E = E_{1} - E_{0}$ between
$\arrowvert 1 \rangle$ and $\arrowvert 0 \rangle$ is shown in 
Fig.\ \ref{transition} as a function of
$B/L$ in a $20 \ nm$ $GaAs/Al_{x}Ga_{1-x}As$ asymmetric quantum dot
($ L= 20 \ nm$).  Several values of Al concentration $x$ are considered.
It is clear from this figure that the transition energy can be tailored
substantially by varying the asymmetry parameter.  With three parameters 
available for adjustment ($B$, $L$, and $x$), we can make $\Delta E$ unique 
for each dot in the register. In this way, we can address a given dot by 
using laser light with the correct photon energy.

The electric field from an electron in one dot shifts the energy
levels of electrons in adjacent dots through electrostatic dipole-dipole
coupling.
By appropriate choice of coordinate systems, the dipole moments
associated with $\arrowvert 0 \rangle$ and
$\arrowvert 1 \rangle$ can be written equal in magnitude but
oppositely directed. The dipole-dipole coupling energy is then
defined as \cite{bib:BarencoDeutsch95}
\begin{equation}
V_{dd}=2 \ \frac{\arrowvert d_{1} \arrowvert \ \arrowvert d_{2} \arrowvert}
{\epsilon_{r} R_{12}^{3}},
\label{Vdd}
\end{equation}
where $d_{1}$ and $d_{2}$ are the ground state dipole moments in the
two dots, $\epsilon_{r} = 12.9$ is the dielectric constant for $GaAs$,
and $R_{12}$ is the distance between the dots.

Figure \ref{coupling} shows the dipole-dipole coupling energy, $V_{dd}$, 
between two asymmetric $GaAs/Al_{x}Ga_{1-x}As$ quantum dots of 
widths $L1 = 19 \ nm$
and $L2 = 21 \ nm$ separated by a $10\ nm$ $Al_{y}Ga_{1-y}As$ barrier.
The coupling energy is plotted as a function of $B/L$ for several values of
$x$ where $B/L$ and $x$ are taken to be the same in both dots. The
dipole-dipole coupling energies are a strongly peaked
function of the asymmetry parameter, $B/L$. From the figure, we see that
values of $V_{dd} \sim 0.15 \ meV$ can be achieved.
By way of comparison, the maximum dipole-dipole coupling energy that can
be achieved with DC biased symmetric quantum dots ($B/L = 0$) is
$V_{dd} = 0.038 \ meV$ at a DC bias field of $F = 112 \ kV/cm$.


Quantum dot electrons can interact with the environment through the
phonon field, particularly the longitudinal-optical (LO) and acoustic (LA) 
phonons.  The LO phonon energy, $\hbar \omega_{LO}$, lies in a narrow
band around $36.2 \ meV$. As long as the quantum
dot energy level spacings lie outside this band, LO phonon scattering
is strongly suppressed by the phonon bottleneck effect. Acoustic
phonon energies are much smaller than the energy difference, $\Delta E$,
between qubit states. Thus acoustic phonon scattering requires
multiple emission processes which are also very slow.  Theoretical
studies on phonon bottleneck effects in GaAs quantum dots indicate that
LO and LA phonon scattering rates including multiple phonon processes
are slower than the spontaneous
emission rate {\em provided that the quantum dot energy level spacing is
greater than $\sim 1$ meV and, at the same time, avoids a narrow window 
(of a few meV) around the LO phonon energy } \cite{bib:Inoshita}.

While dephasing via interactions with the phonon field can be strongly 
suppressed by proper designing of the structure, 
quantum dot electrons are still coupled to
the environment through spontaneous emission and this is the dominant
dephasing mechanism. Decoherence resulting from spontaneous emission
ultimately limits the total time available for a quantum computation
\cite{bib:Ekert96}. Thus, it's important that the spontaneous
emission lifetime be large. The excited state lifetime, $T_{d}$, against
spontaneous emission is \cite{bib:Ekert96}
\begin{equation}
T_{d} = \frac{3 \hbar \ (\hbar c)^{3}}{4 e^{2}\ D^{2}\ \Delta E^{3}} ~ ,
\label{Td}
\end{equation}
where $D = \langle 0 \arrowvert z \arrowvert 1 \rangle$
is the dipole matrix element between $\arrowvert 0 \rangle$
and $\arrowvert 1 \rangle$.

Figure \ref{lifetime} shows the spontaneous emission lifetime of an electron
in qubit state $\arrowvert 1 \rangle$ for a $20 \ nm$ $GaAs/Al_{x}Ga_{1-x}As$
quantum dot as a function of asymmetry parameter, $B/L$, for several
values of Aluminum concentration, $x$. It's immediately obvious from
Fig.\ \ref{lifetime} that the lifetime depends strongly on $B/L$. Depending
on the value of $x$ chosen, the computed lifetime can achieve a maximum
of between 4000 $ns $ and 6000 $ns$.
In general, the maximum lifetime increases with $x$. In Eq.\ (\ref{Td}),
the lifetime is inversely proportional to $\Delta E^{3}$ and $D^{2}$, but
the sharp peak seen in Fig.\ \ref{lifetime} is due \emph{primarily}
to a pronounced minimum in $D$.
In contrast, the spontaneous emission lifetime in a $20 \ nm$
symmetric quantum dot under a DC bias of $F = 112 \ kV/cm$ is only
$1073 \ ns$.


Based on these results, we can estimate parameters for a solid state
quantum register containing a stack of several asymmetric
$GaAs/Al_{0.3}Ga_{0.7}As$ quantum dots in the $L \sim 20 \ nm$ range
separated by $10 \ nm$ $Al_{y}Ga_{1-y}As$ barriers ($y > 0.4$).
An important design goal is obtaining a large
spontaneous emission lifetime and a large dipole-dipole coupling 
energy.  From Figs. \ref{coupling} and \ref{lifetime}, we see that both can
be achieved by selecting an asymmetry parameter, $B/L = 0.8$.
This gives us a spontaneous emission lifetime $T_{d} = 3100 \ ns$ 
and a dipole-dipole coupling energy $V_{dd} = 0.14 \ meV$.
The transition energy between the qubit states
is on the order of $100\ meV$ ($\lambda = 12.4\ \mu m$).
In a quantum computation, the quantum register is optically driven
by a laser as described in Ref. \onlinecite{bib:BarencoDeutsch95}.
In our example, we require a tunable IR laser in the $12\ \mu m$ range
so we can individually address various transitions between coupled
qubit states.

Following initial state preparation, which can be achieved by cooling
the structure to low temperature,
the computation is driven by applying a series of coherent optical
pulses at appropriate intervals.
The $\pi$-pulse duration, $T_{p}$, must be less than the dephasing
time, $T_{d}$ so that many computation steps can be performed
before decoherence sets in. At the same time, the
pulse linewidth must be narrow enough so that we can selectively excite
transitions separated by the dipole-dipole coupling energy, $V_{dd}$.
For transform limited ultrashort pulses, the linewidth-pulsewidth product
is given by the Heisenberg uncertainty principle. Combining these two
restraints, $T_{p}$ must satisfy
\begin{equation}
\frac{\hbar}{2\ V_{dd}}  \ll T_{p} \ll T_{d}.
\label{inequal}
\end{equation} 
Using $V_{dd}$ and $T_{d}$ for our structure,
we obtain $2.4\ ps \ll T_{p} \ll 3.1 \times \ 10^{6}\ ps$.
For highly biased symmetric quantum dots, it is
$8.7 \ ps \ll T_{p} \ll \ 1.1 \times 10^{6}\ ps$ using values
of $V_{dd} = 0.038 \ meV$ and $T_{d} = 1073 \ ns$.
Hence, the number of computational steps that can be executed before
decoherence sets in (i.e., ratio of the upper and lower limits in the 
inequality) is an order of magnitude larger for the proposed
asymmetric structure.


In summary, we have proposed a solid state quantum register based
on a vertically coupled asymmetric single-electron quantum dot structure
that overcomes the problems of weak dipole-dipole coupling and short
decoherence times encountered in earlier quantum dot computing schemes
based on biased symmetric dots.  This structure may provide a realistic 
candidate for quantum computing in solid state systems.

This work was supported, in part, by the Defense Advanced Research Project
Agency and the Office of Naval Research.



\begin{figure}
\caption{Probability density along the confinement direction, $z$, for the
qubit wavefunctions $\arrowvert 0 \rangle$ (solid line)
and $\arrowvert 1 \rangle$ (dot-dashed line).  The inset shows a schematic 
illustration of the conduction bandedge profile in the $ z $ direction.}
\label{density}
\end{figure}

\begin{figure}
\caption{Transition energy, $\Delta E$, between $\arrowvert 0 \rangle$
and $\arrowvert 1 \rangle$ in an $L = 20 \ nm$
$GaAs/Al_{x}Ga_{1-x}As$ asymmetric quantum dot as a
function of $B/L$ for several values of $x$.}
\label{transition}
\end{figure}

\begin{figure}
\caption{Dipole-dipole interaction between two asymmetric
$GaAs/Al_{x}Ga_{1-x}As$ quantum dots of widths $L1 = 19 \ nm$
and $L2 = 21 \ nm$ separated by an $Al_{y}Ga_{1-y}As$ barrier of width
$Wb = 10 \ nm$. The coupling energy is plotted as a function of $B/L$
for several values of $x$. $B/L$ and $x$ are the same for both dots.}
\label{coupling}
\end{figure}

\begin{figure}
\caption{Spontaneous emission lifetime for qubit state
$ \arrowvert 1 \rangle $ in a $GaAs/Al_{x}Ga_{1-x}As$ quantum dot 
with $ L = 20 \ nm$ as a function of $B/L$ for several values of $x$.}
\label{lifetime}
\end{figure}

%
%

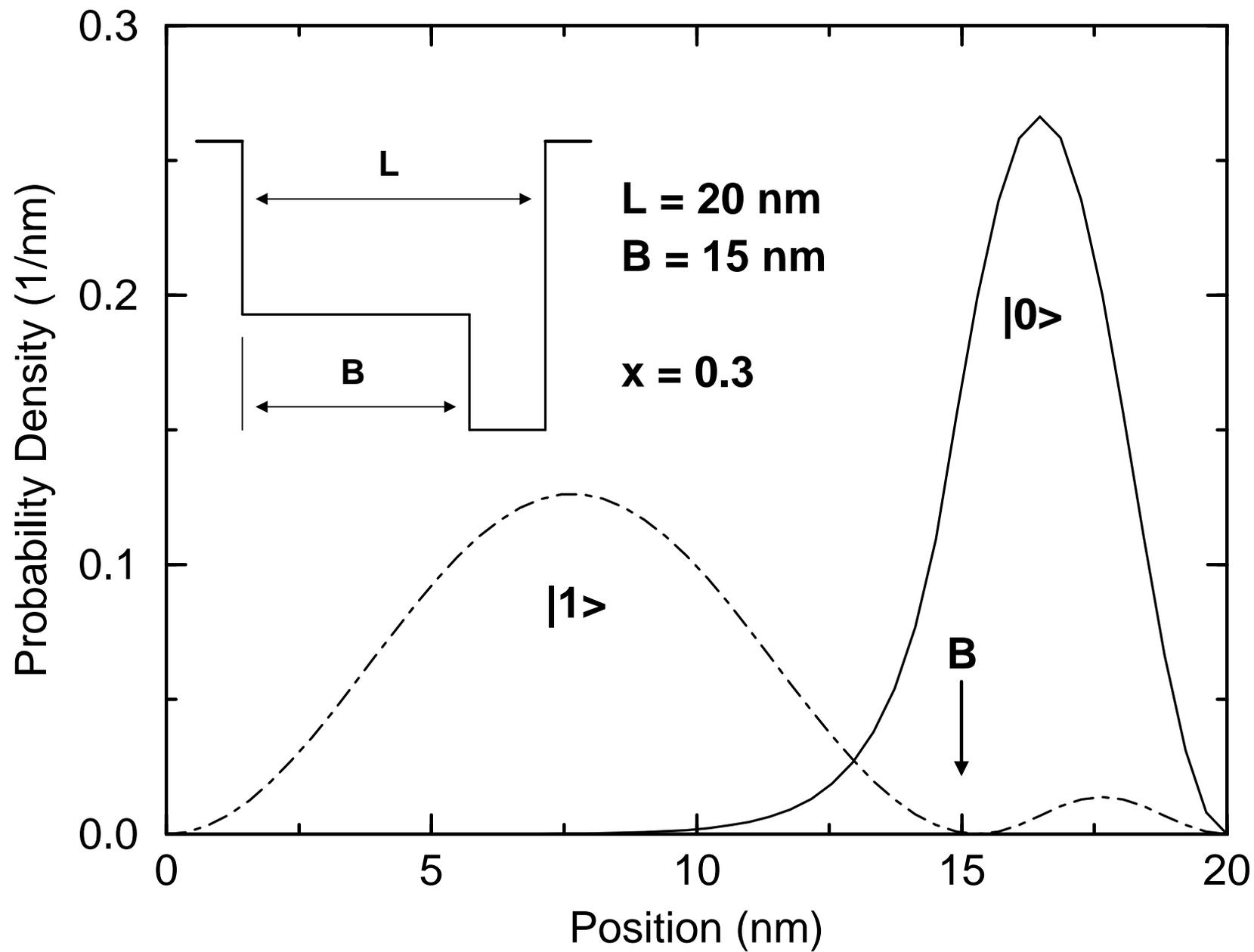
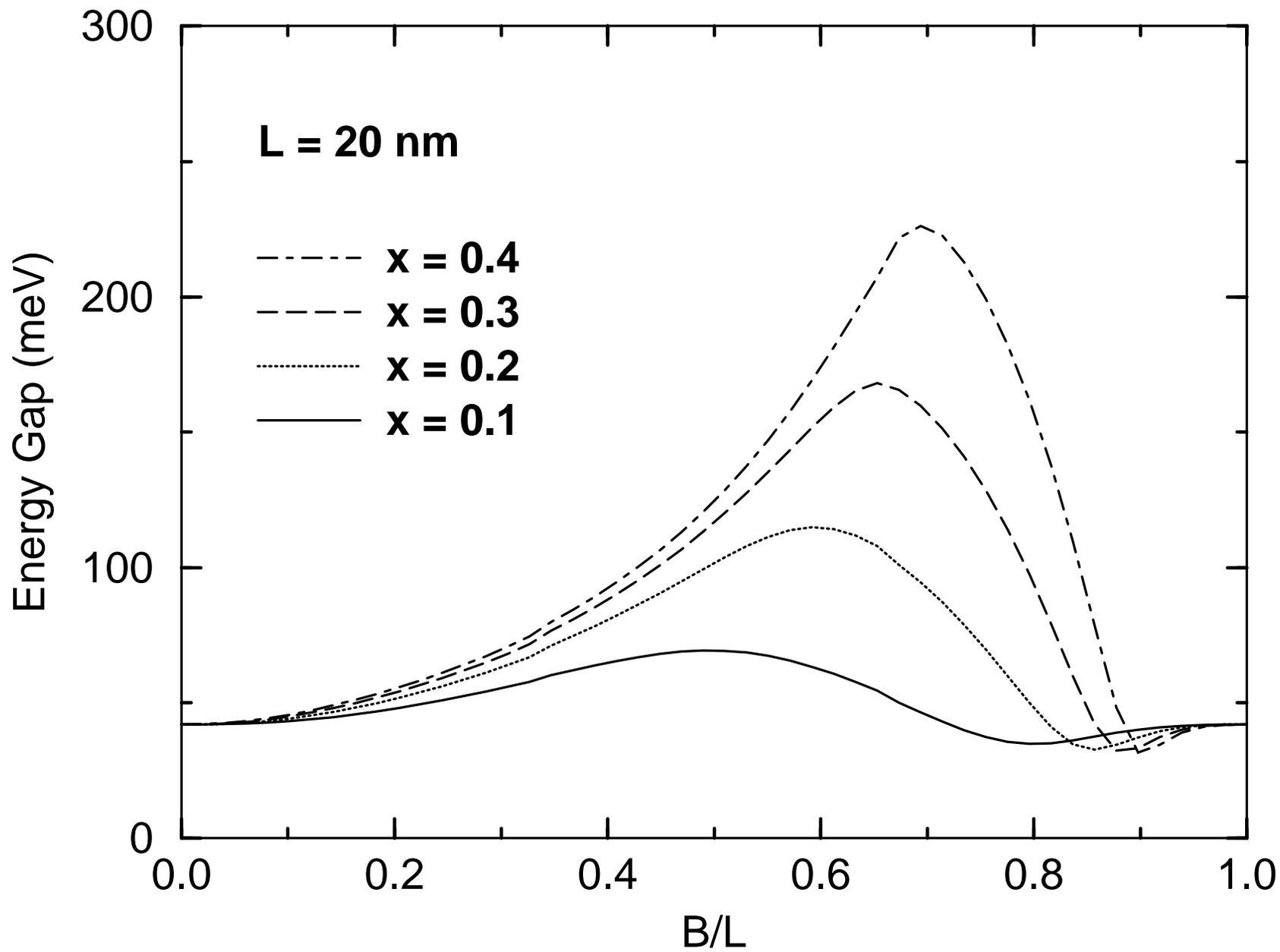
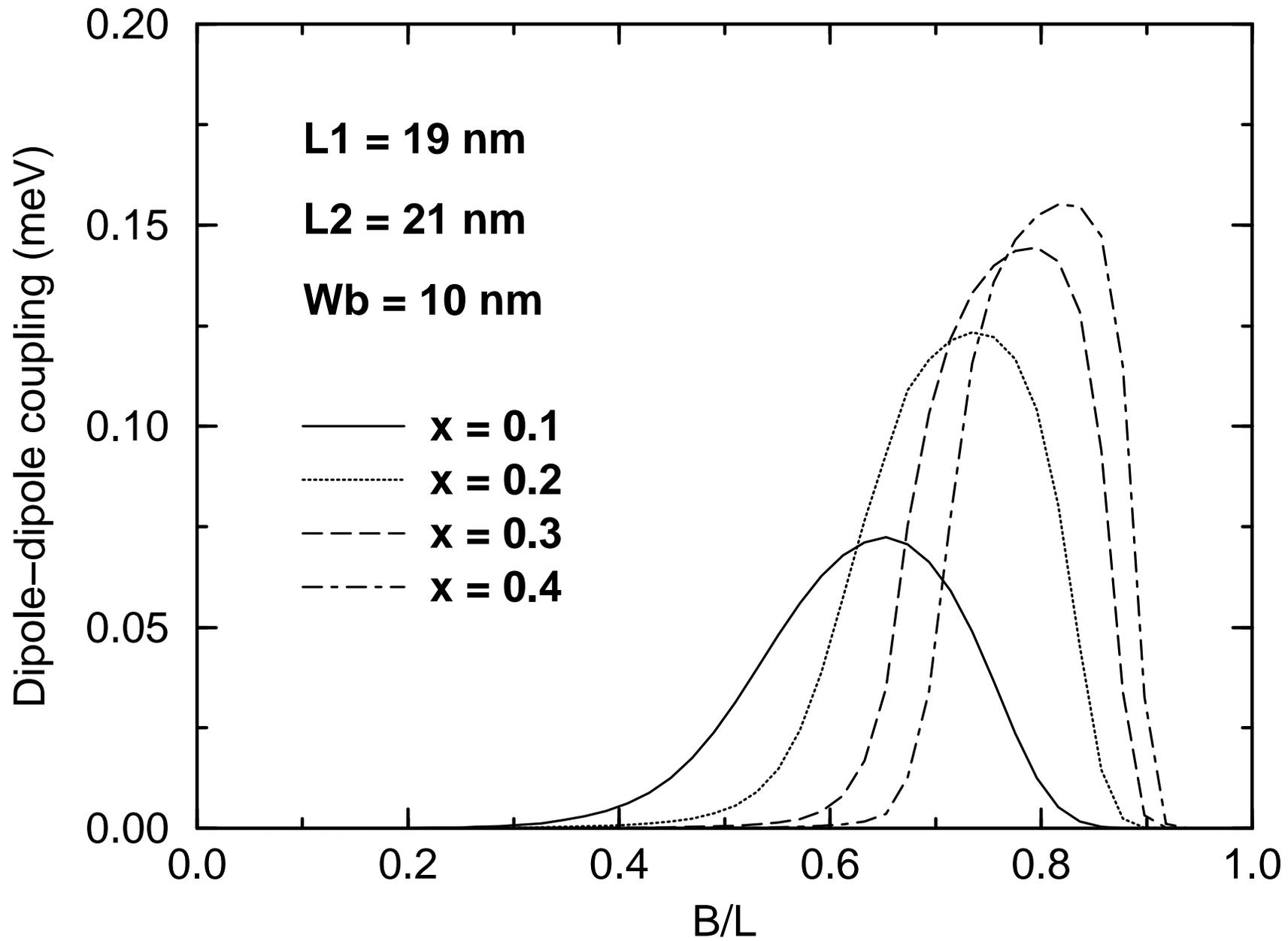
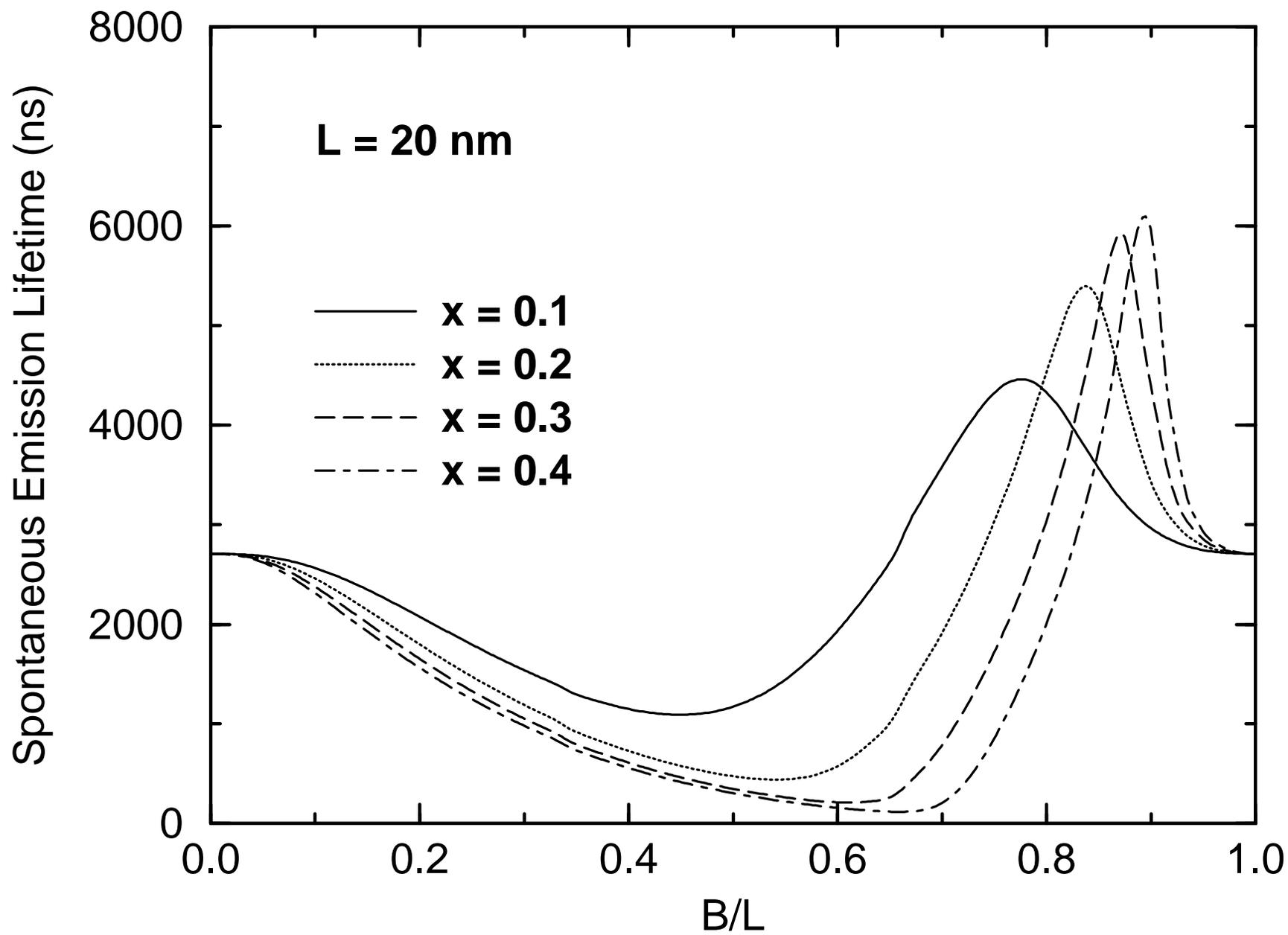
\end{document}